\documentclass[twocolumn,showpacs,preprintnumbers,amsmath,amssymb,aps,prd]{revtex4}
\begin{document}

 \title{\bf Unified First Law and Thermodynamics
 of Dynamical Black Hole in $n$-dimensional Vaidya Spacetime }

 \author{ REN Ji-Rong, LI Ran
 \footnote{Corresponding author. Electronic mail: liran05@lzu.cn}}

 \affiliation{Institute of Theoretical Physics,  Lanzhou University, Lanzhou, 730000, Gansu, China}

 %\date{}

 \begin{abstract}
 {\em\noindent
 As a simple but important example of
 dynamical black hole, we analysis the dynamical black hole
 in $n$-dimensional Vaidya spacetime
 in detail. We investigated the thermodynamics of
 field equation in $n$-dimensional Vaidya spacetime. The unified first law
 was derived in terms of the methods proposed by
 Sean A Hayward. The first law of dynamical black hole
 was obtained by projecting the unified first law along the
 trapping horizon. At last, the
 second law of dynamical black hole is also discussed.}
 \end{abstract}


 \maketitle

 Recently, a remarkable development in black hole theory is the
 proposal of dynamical black hole$^{[1]}$.
 As opposed to the textbook theory of black
 holes, which mostly concerns either stationary
 space-times or event horizons, dynamical black
 hole can be defined locally without the knowledge of the whole
 space-time, which will be bring important applications to
 numerical relativity.

 The vaidya metric has drawn a lot of attention
 as a simplified example of dynamical black hole$^{[2]}$.
 The main purpose of this paper is to
 investigate the unified first law and thermodynamics
 of dynamical black hole in $n$-dimensional Vaidya spacetime.
 We have found the unified first law for $n$-dimensional Vaidya spacetime where the
 work term vanishes naturally because the stress energy describes
 the pressureless fluid. The first and second law of
 thermodynamics is also discussed from which the entropy is given
 by $S=A/4$. This coincides with the previous result$^{[3]}$.
 The method used in this letter has been developed to discuss the
 thermodynamics of apparent horizon in FRW universe in different
 gravity model$^{[4]}$.

 The key geometrical objects are dynamical horizons,
 which are hypersurfaces $H$ (in space-time) foliated by
 marginal surfaces. A marginal surface is a spatial surface on
 which one null expansion vanishes, where the null expansions $\theta_\pm$
 may be defined by
 \begin{equation}
  \theta_\pm=\frac{L_\pm \delta A}{\delta A}\;\;,
 \end{equation}
 where $\delta A$ denotes the area
 element and $L_\pm$ the Lie derivatives along future-pointing null
 normal vectors $l_\pm$. This is a surface where outgoing or ingoing
 light rays are just trapped, neither converging nor diverging.

 When the marginal surface has the topology of $(n-2)$-sphere,
 the null expansions can be explicitly expressed as
 \begin{equation}
 \theta_\pm=(n-2)\frac{\partial_\pm r}{r}\;\;,
 \end{equation}
 where $r$ is used to label the marginal surface.
 $H$ is generated by the vector $\partial/\partial r$, normal to
 the marginal surfaces.

 For the author's local definition$^{[1]}$ of black hole as a future ($\theta_-<0$
 on $H$, for $\theta_+=0$) outer ($L_-\theta_+<0$) trapping
 horizon. If Einstein gravity and dominant energy condition are
 assumed, it is shown that the area of marginal surface does'nt
 decrease which is similar to Hawking's so-called
 second law for event horizons, but for a physically locatable
 horizon. The more detailed discussions can be found in
 Ref[1].

 Now, we will derive the unified first law for
 Vaidya spacetime to make a clear identification
 to the method proposed by Sean A Hayward. The definition of some
 quantities strictly follows Ref[5].

 The $n-$dimensional $(n>3)$ Vaidya spacetime$^{[6]}$ is described by the metric
 \begin{equation}
 ds^2=-(1-\frac{2m(v)}{(n-3)r^{n-3}})dv^2+2dvdr+r^2d\Omega^2_{n-2}\;\;.
 \end{equation}
 where $d\Omega^2_{n-2}$ stands for the metric of the
 unity $(n-2)$-dimensional sphere.

 The stress energy tensor is determined by the derivative of $m(v)$
 \begin{equation}
 T_{\mu\nu}=\frac{1}{8\pi}\frac{(n-2)}{(n-3)r^{n-2}}\frac{dm(v)}{dv}l_\mu l_\nu\;\;,
 \end{equation}
 where $l_\mu=-\partial_\mu v$ is the tangent vector of ingoing null
 geodesics. The mass function $m(v)$ must increase to satisfy the
 energy condition.

 This is an astrophysically unrealistic
 toy model, but it does serve as a very useful testing ground
 for dynamical black hole in numerical relativity$^{[2]}$. It
 has been extensively used, for example, to study the formation of
 naked singularities$^{[6]}$.

 The dual-null coordinate can be introduced as
 \begin{eqnarray}
 &&d\xi^+=\frac{1}{2}dv\;\;\\
 &&d\xi^-=(1-\frac{2m(v)}{(n-3)r^{n-3}})dv-2dr\;\;
 \end{eqnarray}
 Then, the Vaidya metric can be put into the dual-null form
 \begin{equation}
 ds^2=-2d\xi^+ d\xi^- +r^2d\Omega^2_{n-2}\;\;.
 \end{equation}
 It is easy to find the relation
 \begin{eqnarray}
 &&\frac{\partial}{\partial\xi^+}=2\frac{\partial}{\partial
 v}+(1-\frac{2m(v)}{(n-3)r^{n-3}})\frac{\partial}{\partial r}\;\;,\\
 &&\frac{\partial}{\partial\xi^-}=-\frac{1}{2}\frac{\partial}{\partial r}\;\;.
 \end{eqnarray}

 The null expansions is given by
 \begin{eqnarray}
 &&\theta_+=(n-2)(\frac{1}{r}-\frac{2m(v)}{(n-3)r^{n-2}})\;\;,\\
 &&\theta_-=-\frac{(n-2)}{2r}\;\;,
 \end{eqnarray}
 from which one can see that the dynamical horizon is located at
 $r_{\textrm{DH}}=(2m(v)/(n-3))^{1/(n-3)}$, i.e. the null expansion $\theta_+=0$.

 One can also find
 \begin{equation}
 \partial_-\theta^+=-\frac{(n-2)}{2r^2}(\frac{(n-2)}{(n-3)}\frac{2m(v)}{r^{n-3}}-1)\;\;.
 \end{equation}
 So, on the dynamical horizon $r_{\textrm{DH}}$, one have
 \begin{equation}
  \partial_-\theta^+=-\frac{(n-2)(n-3)}{2r_{\textrm{DH}}}<0\;\;.
 \end{equation}
 The dynamical horizon can be expressed as
 \begin{equation}
 \Phi=r-r_{\textrm{DH}}=0\;\;,
 \end{equation}
 from which one can calculate that
 \begin{eqnarray}
 g^{\mu\nu}\partial_\mu\Phi\partial_\nu\Phi
 &=&-4(\frac{1}{n-3})^{(\frac{n-2}{n-3})}\nonumber\\
 &&\times(2m(v))^{-(\frac{n-4}{n-3})}\frac{dm(v)}{dv}\;\;.
 \end{eqnarray}
 So the dynamical horizon here is a spacelike hypersurface since
 $dm/dv>0$(so that the energy condition is satisfied). In vaidya
 spacetime, the dynamical horizon is not consistent with the event horizon
 because event horizon must be a null hypersurface from the
 definition of event horizon.

 The Misner-Sharp energy$^{[7-9]}$ is defined as
 \begin{equation}
 E=\frac{(n-1)(n-2)V_{n-1}}{16\pi}r^{n-3}(1-\nabla^a r\nabla_a
 r)\;\;,
 \end{equation}
 where $a$ is the index of transverse space orthogonal to the
 tangent space of marginal surface, i.e. $a=\{+,-\}$.
 $V_{n-1}=\pi^{(\frac{n-1}{2})}/\Gamma(\frac{n+1}{2})$
 denotes the volume of the $(n-1)$ dimensional unit ball.

 The Misner-Sharp energy at $r=r_{\textrm{DH}}$ is just the mass of
 dynamical black hole in Vaidya spacetime.
 Various properties of the Misner-Sharp energy in the spherically symmetric
 spacetime is established by Sean A. Hayward in Ref[8]. This energy is the
 total energy inside the sphere with radius $r$. The important
 advantage of Misner-Sharp energy which has the relation to the field
 equation will be shown below.

 In the dual-null coordinate, the non-vanishing component of
 stress energy tensor is
 \begin{equation}
 T_{++}=\frac{1}{2\pi}\frac{(n-2)}{(n-3)r^{n-2}}\frac{dm(v)}{dv}\;\;.
 \end{equation}

 According to the dual-null form of Vaidya metric,
 Einstein equation $G_{\mu\nu}=8\pi T_{\mu\nu}$ can be written as
 \begin{eqnarray}
 &&\partial_+\partial_+ r=-\frac{4}{(n-3)}\frac{1}{r^{n-3}}\frac{dm}{dv}\;\;,\\
 &&\partial_-\partial_- r=0\;\;,\\
 &&(n-3)\partial_+r\partial_-r+r\partial_+\partial_-r+\frac{1}{2}(n-3)=0\;\;.
 \end{eqnarray}

 Energy density $\omega$ is defined as
 \begin{eqnarray}
 \omega=-\frac{1}{2}\textrm{Tr}\;T\;\;,
 \end{eqnarray}
 where the trace is performed also in the transverse space orthogonal to the
 tangent space of marginal surface.
 It is easy to calculate that $\omega$ vanishes in the Vaidya
 spacetime.

 Energy flux or momentum density $\psi$ is defined as
 \begin{eqnarray}
 \psi=T\nabla r+\omega\nabla r\;\;.
 \end{eqnarray}
 A  direct calculation shows that
 \begin{eqnarray}
 &&\psi_+=-\frac{1}{2\pi}\frac{(n-2)}{(n-3)r^{n-2}}\frac{dm}{dv}\partial_-r\;,\\
 &&\psi_-=0\;\;.
 \end{eqnarray}

 From the definition of Misner-Sharp energy, using
 Einstein equation and the definition of energy flux,
 one can show that
 \begin{eqnarray}
 &&\partial_+E=A\psi_+\;,\\
 &&\partial_-E=0\;\;,
 \end{eqnarray}
 where $A=A_{n-2}r^{n-2}$ is the area of marginal $(n-2)$-sphere
 with $r$ being the radius and $A_{n-2}=(n-1)V_{n-1}$ being
 the area of the $(n-2)$-dimensional sphere.

 The above two equations can be uniformly written as
 \begin{equation}
 dE=A\psi\;\;.
 \end{equation}
 This is just the unified first law in Vaidya spacetime.
 Because the stress energy $T_{\mu\nu}$ describes the null dust, a
 pressureless fluid with energy density $\frac{1}{8\pi}\frac{(n-2)}{(n-3)r^{n-2}}\frac{dm(v)}{dv}$ and
 velocity $l_\mu$, the work term vanishes naturally and only
 the energy supply term is remained.

 Now, we turn to the discussion of the first law of dynamical
 black hole in Vaudya spacetime.
 The dynamical surface gravity is defined as
 \begin{equation}
 \kappa=\frac{1}{2}\nabla^a\nabla_a r\;\;.
 \end{equation}
 For Vaidya spacetime, using the Einstein equation one can obtain
 \begin{equation}
 \kappa=\frac{1}{2r_{\textrm{DH}}}\;\;.
 \end{equation}

 Introduce a vector $z$ that is tangent to the dynamical horizon.
 In the dual-null coordinate, $z$ can be expressed as
 \begin{equation}
 z=z^+\partial_++z^-\partial_-\;\;.
 \end{equation}
 Since $\partial_+ r$ always vanishes on the dynamical horizon,
 the projection of $\partial_+ r$ to the dynamical horizon also
 vanishes. So, we have
 \begin{equation}
 z^+\partial_+\partial_+r+z^-\partial_-\partial_+r=0\;\;.
 \end{equation}

 One can obtain the first law of dynamical black hole through
 projecting the unified first law to the dynamical horizon. The
 key procedure is to prove the relation
 \begin{equation}
 \langle A\psi,z\rangle=\frac{\kappa}{8\pi}\langle
 dA,z\rangle\;\;.
 \end{equation}
 where $\langle\; ,\;\rangle$ denotes the inner product.

 In our situation, this relation is indeed easy to prove. From the
 above relation, if we identify the temperature $T$ of dynamical
 black hole with $T=\frac{\kappa}{2\pi}$ and the left hand side of
 above equation with $\delta Q$, one can get the Clausius relation
 \begin{equation}
 \delta Q=TdS\;\;,
 \end{equation}
 where $S=\frac{A}{4}$ is the entropy of dynamical black hole
 with $A=A_{n-2}r_{\textrm{DH}}^{n-2}$ being the area of dynamical horizon.

 Ted Jacobson$^{[10]}$ have been able to derive the Einstein
 equation from the proportionality of entropy
 and horizon area together with the Clausius relation $\delta Q=TdS$
 connecting heat, entropy, and temperature. Recently, further
 research$^{[11]}$ dedicates that it is impossible to derive the field
 equation of $f(R)$ gravity from the Clausius relation $\delta
 Q=TdS$ in terms of the viewpoint of equilibrium thermodynamics.
 This problem can be canceled by adding a entropy production term
 $d_iS$ to the Clausius relation which is ultimately changed into
 $dS=\frac{\delta Q}{T}+d_i S$. Now, we have found the thermodynamics
 intrinsic in the Einstein equation. This implies that there will
 be deep relationship between the thermodynamics and gravity
 theory.

 By projecting the unified first law to the
 horizon, the first law of dynamical black hole thermodynamics
 can be obtained
 \begin{equation}
 \langle dE,z\rangle=\frac{\kappa}{8\pi}\langle dA,z\rangle\;\;.
 \end{equation}
 This equation is straightforward from the equation (27) and (32).

 At last, we briefly discuss the second law of dynamical black
 hole in Vaidya spacetime. As have been discussed, the area of
 dynamical horizon is $A=A_{n-2}(\frac{2m(v)}{n-3})^{\frac{n-2}{n-3}}$.
 Since the mass $m(v)$ function
 increase to satisfy the energy condition, the area $A$ of
 dynamical black hole can not decrease, i.e. the entropy of
 dynamical black hole must increase. This is just the second law
 of dynamical black hole thermodynamics.

 In conclusion, we have discussed the thermodynamics of dynamical
 black hole in $n-$dimensional Vaidya spacetime. In this paper, Einstein' gravity is
 assumed. If the unified first law is also valid for other gravity theory
 will be studied in the future.

 \emph{ The author LI Ran thanks Dr.Li-Ming Cao
 for pointing out some defects in our original manuscript and
 bringing our attention to Reference} [4].

 \vspace*{0.2truecm}
 \def\REF#1{\par\hangindent\parindent\indent\llap{#1\enspace}\ignorespaces}

 \section*{\Large\bf References}
 \vspace*{-0.8\baselineskip}
 \hskip 7pt {\footnotesize

 \REF{[1]} S. A. Hayward, \textit{Phys. Rev. Lett.} \textbf{93} (2004) 251101
 \REF{} A. Ashtekar and B. Krishnan, \textit{Phys. Rev. Lett.} 89 (2002) 261101
 \REF{} A. Ashtekar and B. Krishnan, \textit{Living Rev. Rel.} \textbf{7} (2004) 10

 \REF{[2]} E. Schnetter and B. Krishnan, \textit{Phys. Rev.} D \textbf{73} (2006) 021502
 \REF{} S. Sawayama, \textit{Phys. Rev.} D \textbf{73} (2006) 064024
 \REF{} Ishai Ben-Dov, \textit{Phys. Rev.} D \textbf{75} (2007) 064007

 \REF{[3]} S. A. Hayward, S. Mukohyama and M. C. Ashworth, \textit{Phys. Lett.} A \textbf{256} (1999) 347

 \REF{[4]} Rong-Gen Cai and Li-Ming Cao \textit{Phys.Rev.} \textbf{D} 75 (2007) 064008

 \REF{[5]} S. A. Hayward, \textit{Class. Quantum Grav.} \textbf{15} (1998) 3147

 \REF{[6]} S. G. Ghosh and N. Dadhich, \textit{Phys. Rev.} D \textbf{64} (2001) 047501
 \REF{} S. G. Ghosh and N. Dadhich, \textit{Phys. Rev.} D \textbf{65} (2002) 127502

 \REF{[7]} C. W. Misner and D. H. Sharp, \textit{ Phys. Rev.} \textbf{136} (1964) B571

 \REF{[8]} S. A Hayward, \textit{ Phys. Rev.} D \textbf{53} (1996) 1938

 \REF{[9]} D. Bak and S. J. Rey, \textit{Class. Quantum Grav.} \textbf{17}(2000)L83

 \REF{[10]} T. Jacobson, \textit{ Phys. Rev. Lett.} \textbf{75} (1995) 1260

 \REF{[11]} C. Eling, R. Guedens and T. Jacobson, \textit{ Phys. Rev. Lett.} \textbf{96} (2006) 121301

 \end{document}